\documentstyle[11pt,ysc,twoside,epsf]{article}
\def\edcomment#1{\iffalse\marginpar{\raggedright\sl#1\/}\else\relax\fi}
\reversemarginpar

\begin{document}
\title{Optical parameters of the nonisothermal Uranus's and Neptune's atmospheres}

\author{N.M.Kostogryz}
\affil{Main Astronomical Observatory of NAS of Ukraine,
27,Zabolotnoho str., Kyiv, Ukraine, 03680}

\begin{abstract}

A method of the calculation of optical parameters of the
nonisothermal giant planet atmospheres was developed using
detailed intensity data of Raman scattering. We have used the
model of Morozhenko (A.V. Morozhenko, 1997) as a baseline. In such
a way, using observational data of Uranus and Neptune
(E.Karkoschka, 1994), the spectral values of ratio of optical
depth components: aerosol and gas components $\tau_a/\tau_R$,
absorbing and scattering components $\tau_\kappa/\tau_R$, and also
single scattering albedo of aerosol component corrected for Raman
scattering $\omega'$ were obtained (where $\tau_a, \tau_R$ are
aerosol and gas components, and $\tau_\kappa$ is absorbing
components of effective optical depths of the formation of
diffusely reflected irradiation). The averaged value of ratio
$\tau_a/\tau_R$ is 0.96 but it slowly decreases in the spectral
range of 350-450nm for Uranus and $\tau_a/\tau_R$ is 1.35 for
Neptune.
\end{abstract}

\section{Introduction}

The atmospheres of Uranus and Neptune are known to be composed
predominantly of molecular hydrogen. Since there is so much $H_2$
in the atmospheres of outer planets and $H_2$ has a reasonably
strong Raman spectrum, it is very important to attempt to
understand the physics of planetary Raman scattering.

Raman scattering is the incoherent non-resonance scattering of
photons by a molecule. During molecular scattering process, the
photon may loose energy according to certain molecular
transitions. If the incident solar photon of frequency $\nu_0$ is
scattered, it will emerge at frequency $\nu_0 \pm
\bigtriangleup\nu$, where $\bigtriangleup\nu$ is the frequency of
the Raman transition of the molecule.

In recent years the observational data of detailed intensity of
Raman scattering in the giant planet spectra was proposed to use
to determine the relative contribution of the aerosol component of
atmosphere. In such a way, we can determine the values of aerosol
to gas ratio of optical depth components $\tau_a/\tau_R$, and
absorbing to scattering ratio $\tau_\kappa/\tau_R$
(M.S.Dementiev,1992; A.V.Morozhenko, 1997). In these papers, the
model of atmosphere was taken to be isothermal, while the real
giant planet atmospheres have complex temperature profiles
(G.F.Lindal, et al., 1987; G.F.Lindal, et al., 1990). The relative
number of hydrogen molecules in the ortho- and para- state depends
on the depth in the nonisothermal atmosphere, while it doesn't
depend on the depth in the isothermal one. So, the detailed
intensity of Raman scattering will depend on the effective optical
depth of the formation of diffusely reflected irradiation. The
method of accounting of the real temperature profile in computing
of Raman scattering effects was developed by Morozhenko and
Kostogryz (A.V.Morozhenko and N.Kostogryz, 2005).

This paper presents a method of computation of optical parameters
of the Uranus's and Neptune's atmospheres such as $\tau_a/\tau_R$
and $\tau_\kappa/\tau_R$ considering detailed intensity of Raman
scattering and using observational data of Uranus's and Neptune's
atmospheres (E.Karkoschka, 1994) and experimental temperature
profiles (G.F.Lindal, et al., 1987; G.F.Lindal, et al., 1990).

Section 2 contains reviews of the model of atmosphere. Section 3
is devoted to the method of computation and sections 4 and 5
describe results of computation and some conclusions of this work.

\section{Model of atmosphere}
 As reliable information about vertical structure is lacking,
 especially about aerosol component of atmosphere, we will use
 the model of homogeneous semi-infinite gas-aerosol layer. We
 take into account that gas components of atmosphere are
 hydrogen(85$\%$) and helium(15$\%$), and also we consider
 nonisothermal atmospheres of Uranus and Neptune using experimental
 temperature profiles (G.F.Lindal, et al., 1987; G.F.Lindal, et al., 1990).

 Raman scattering is considered for the four major hydrogen
 transitions, the rotational S(0), S(1) and O(2) and the
 vibrational $Q_1(1)$ transitions, which produce significant Raman
 ghosts. Raman shifts and cross sections at 400nm of several
 transitions of molecules of interest in planetary atmospheres
 were taken from Cochran and Trafton(W.D.Cochran and L.M.Trafton, 1978).

 At the temperatures of planetary atmospheres, most of the
 molecules are in the lower rotational levels of the ground
 vibrational state. Therefore the Stocks component of the
 Raman scattering will dominate. The molecule will absorb energy
 and the photon will emerge at lower frequency or longer
 wavelength than it entered.

 The estimation of the spectral values of single scattering albedo
 was obtained via comparison of observational data of geometric
 albedo (E.Karkoschka, 1994) and theoretical computed for homogeneous semi-infinite layer
 with a Rayleigh scattering indecatrix by Ovsak(A.V.Morozhenko, 2004).

 The values of effective pressure on that intensity of diffuse
 reflected radiation is forming were taken from Morozhenko's paper (A.V.Morozhenko, 2006)

\section{Method of computation}

To take into account Raman scattering effects equation of single
scattering albedo from Pollack was used (J.B.Pollack et al.,
1986). While Pollack considers only photons which are shifted to
$\lambda_0$ due to Raman scattering, we propose more correct
expression for single scattering albedo, which account still
photons which are shifted from $\lambda_0$ due to Raman
scattering.

\begin{equation}
\omega=\frac{\tau_a/\tau_R +
D}{1+\tau_a/\tau_R+\tau_\kappa/\tau_R} \label{omega}
\end{equation}

\begin{equation}
D=1+0.85*[((N_0\tau_{S(0)}+N_2\tau_{O(2)})f_{{\lambda}_1}+N_1\tau_{S(1)}f_{{\lambda}_2}+\tau_{Q_1(1)}f_{{\lambda}_3})/f_{{\lambda}_0}\tau_R]-A
\label{parD}
\end{equation}

\begin{equation}
A=0.85*(N_0\tau_{S(0)}+N_2\tau_{O(2)}+N_1\tau_{S(1)}+\tau_{Q_1(1)})/\tau_R
\end{equation}

\noindent where $f_{{\lambda}_1}, f_{{\lambda}_2},
f_{{\lambda}_3}$ are spectral values of the energy in Solar
spectrum on the wavelength, from which rotation $(S(0), O(2),
S(1))$ and vibration $(Q_1(1))$ Stocks transition of Raman
scattering carries the sun photon on wavelength $\lambda_0$
accordingly; $\tau_{S(0)}, \tau_{O(2)}, \tau_{S(1)} $ and
$\tau_{Q_1(1)}$ are optical depths of the Raman scattering of
corresponding transition.

The amount of hydrogen molecules in ortho- and para- state
described as

\begin{equation}
N=3(2j+1)\exp{-B*j(j+1)\frac{hc}{kT}};
 j=1,3,5\ldots
 \end{equation}
\begin{equation}
N=(2j+1)\exp{-B*j(j+1)\frac{hc}{kT}}; j=0,2,4\ldots
 \end{equation}

\noindent where B is rotation constant, which for hydrogen molecule
is equal $60 cm^{-1}$, j is quantum number, h is Planck constant, c
is speed of light, k is Boltzman constant, T is temperature (in
Kelvin).

Solar spectrum (L.Delbouille, et al., 1973) which reduced to 1-nm
resolution was used in our investigations.

It was the first step of our code which was described in details
by Morozhenko and Kostogryz (A.V. Morozhenko, N. Kostogryz, 2005)

The estimation of parameter D was obtained using (2) and the
spectral values of single scattering albedo were obtained via
comparison of observational data of geometric  albedo
(E.Karkoschka, 1994) and computed for homogeneous semi-infinitive
layer with Rayleigh scattering indecatrix by Ovsak
(A.V.Morozhenko, 2004). Then we approximate that ($\tau_a/\tau_R$)
is constant every 10 nm from 350-450 nm. Using min-square method,
we obtained such a value of ($\tau_\kappa/\tau_S$) where
$\tau_\kappa/\tau_S = (\tau_\kappa/\tau_R)/(1+\tau_a/\tau_R)$ has
minimum dispersion. At that, ($\tau_a/\tau_R$) can have values
from 0 to 3 with 0.03 resolution. After that, we average all
values of ($\tau_a/\tau_R$) over spectral interval 350-450 nm on
which methane bands are practically absent and find spectral
dependence of ($\tau_\kappa/\tau_S$).

After determination of ($\tau_a/\tau_R$), single scattering albedo
($\omega'$) was corrected for Raman Scattering using such a
formula:

\begin{equation}
1/\omega'=(D+\tau_a/\tau_R)/((1+\tau_a/\tau_R)\ast\omega)
 \end{equation}

\section{Results and discussions}
A method of accounting complex temperature profile for determine
such parameters of atmosphere as ($\tau_a/\tau_R$),
($\tau_\kappa/\tau_R$) and ($\tau_\kappa/\tau_S$) was developed
and the averaged value of ratio of the optical depth component:
aerosol and gas component ($\tau_a/\tau_R=0.96$) for Uranus, and
($\tau_a/\tau_R=1.35$) for Neptune were obtained and the spectral
dependence of ($\tau_\kappa/\tau_S$) was also obtained.

\begin{figure}
\plotone{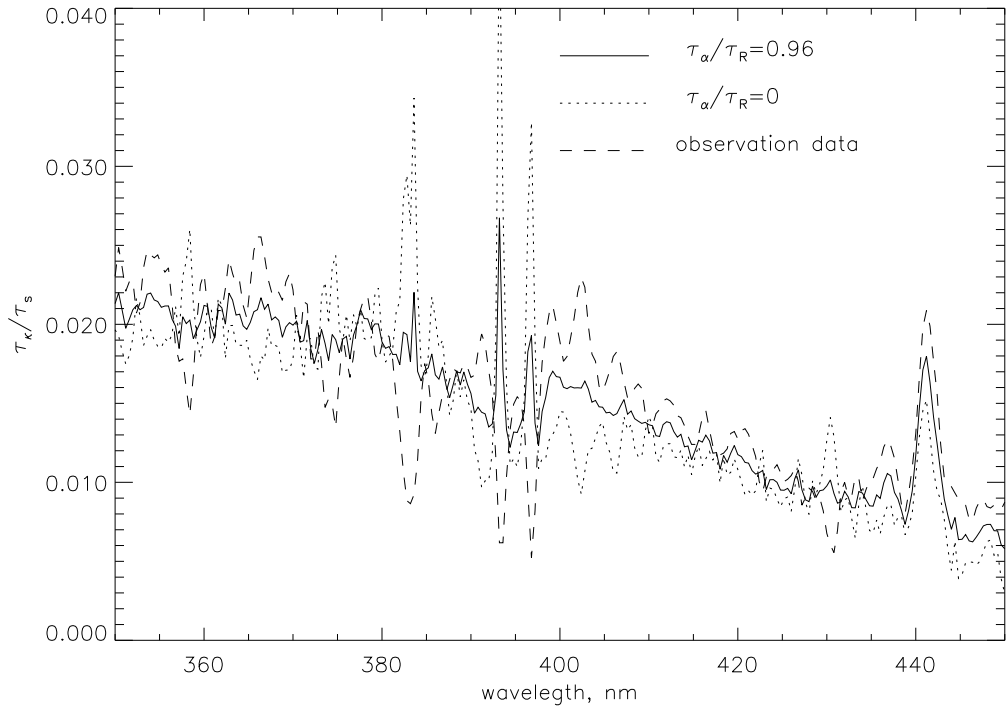} \caption{Spectral dependence of
($\tau_\kappa/\tau_S$) for Uranus. Observational data denoted as
dotted line, points are for gas atmosphere ($\tau_a/\tau_R=0$),
and solid line is for nonisothermal gas-aerosol atmosphere
($\tau_a/\tau_R=0.96$)}
\end{figure}

In fig.1 comparison of spectral dependence of
($\tau_\kappa/\tau_S$) for observational data of Uranus which
determined as $\tau_\kappa/\tau_S = 1/\omega-1$, for gas
atmosphere ($\tau_a/\tau_R=0$), and for nonisothermal gas-aerosol
atmosphere ($\tau_a/\tau_R=0.96$) is depicted. It is shown, that
the values of ($\tau_\kappa/\tau_S$) have minimum dispersion at
($\tau_a/\tau_R=0.96$). So, we can obtained ($\tau_\kappa/\tau_S$)
corrected for Raman scattering.

\begin{figure}
\plotone{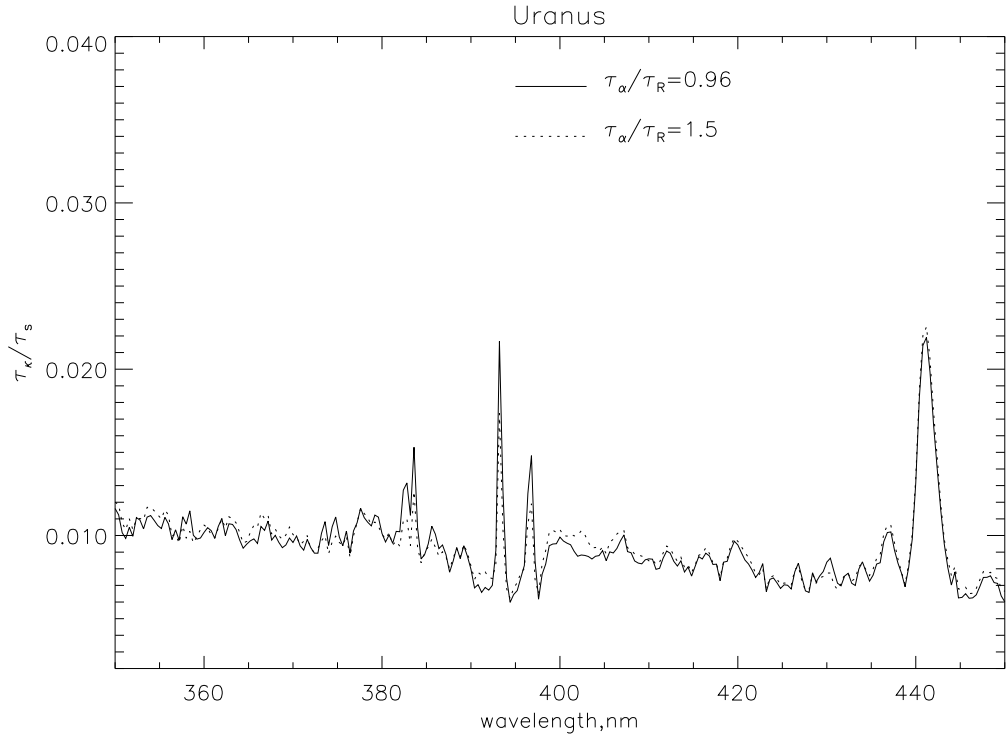} \caption{Spectral dependence of
($\tau_\kappa/\tau_S$) for isothermal(dotted line) and
nonisothermal(solid line) Uranus atmosphere}
\end{figure}

Morozhenko and Kostogryz(A.V.Morozhenko and N.Kostogryz, 2005)
showed that ignoring of real temperature profile could lead to large
errors in determination of ($\tau_a/\tau_R$), ($\tau_\kappa/\tau_R$)
and ($\tau_\kappa/\tau_S$). In this paper we check this result using
real values of these parameters. For this purpose we determined all
these parameters for isothermal atmosphere with effective
temperature $T = 57.1K$ and  such value ($\tau_a/\tau_R =
1.5$)(fig.2) was obtained.

\begin{figure}
\plotone{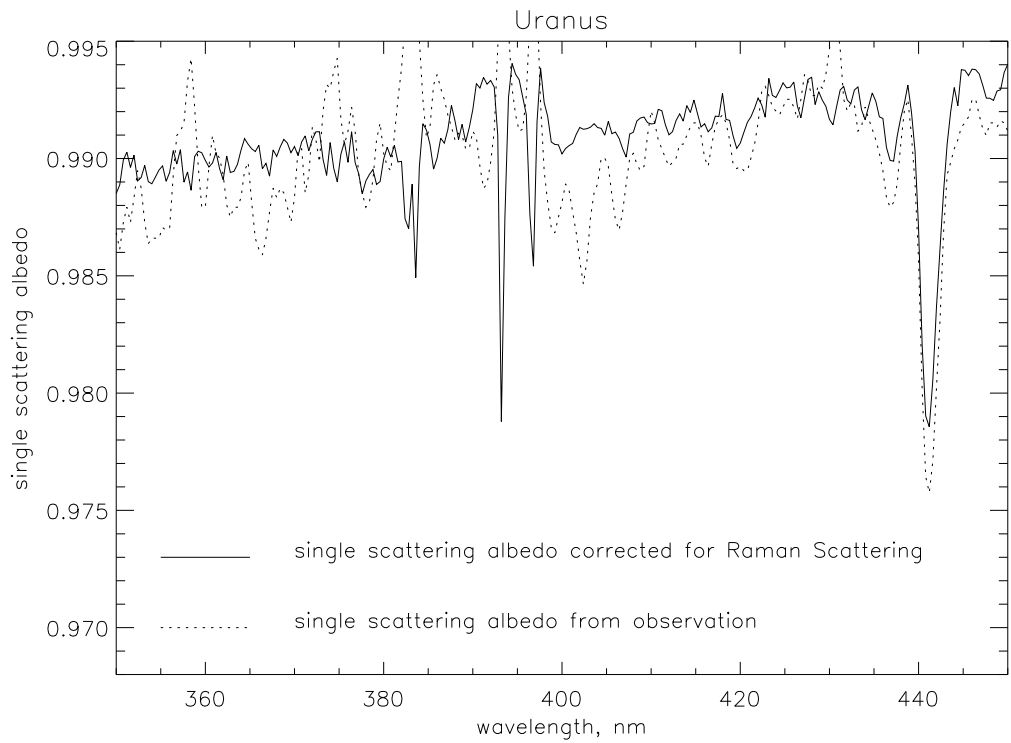} \caption{Spectral dependence of single
scattering albedo for Uranus. Observational data denoted as dotted
line, and solid line is single scattering albedo corrected for
Raman scattering}
\end{figure}

It is known, that one of the Raman scattering effects in the
Uranus's and Neptune's atmospheres is decreasing of geometric
albedo and single scattering albedo in the UV spectra that can be
discussed as pseudo-absorption in continuum. Using this method, we
computed values of single scattering albedo corrected for Raman
scattering and compared them with observed ones (fig.3).

\begin{figure}
\plotone{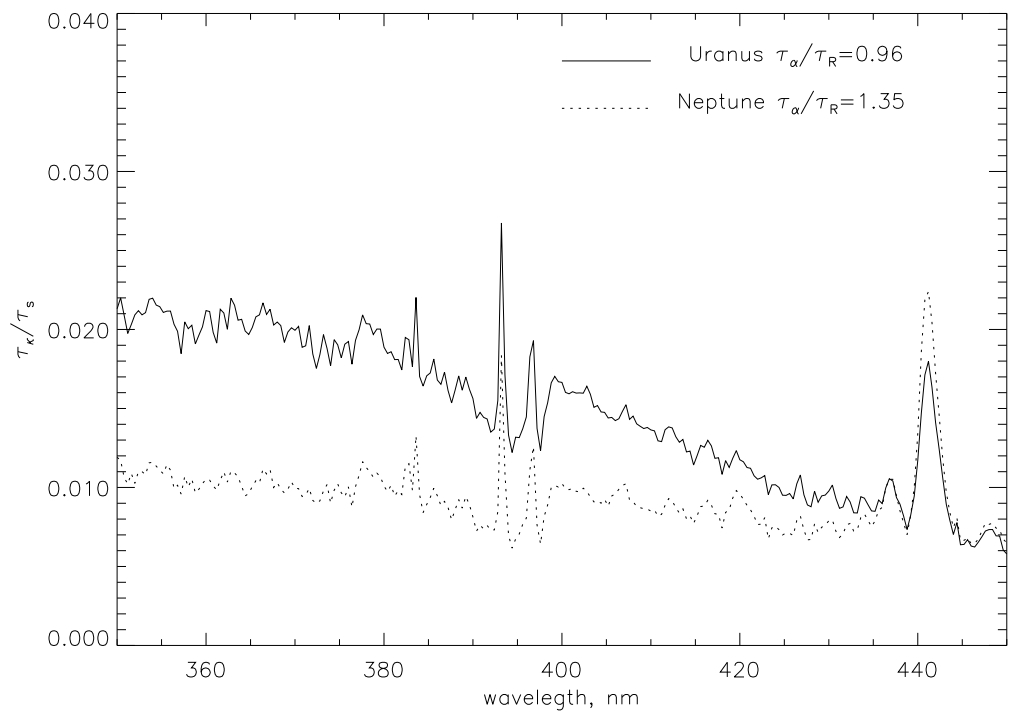} \caption{Spectral dependence of
($\tau_\kappa/\tau_S$) for Uranus(firm line) and Neptune(dotted
line). }
\end{figure}

The values of aerosol and gas ratio of optical depth components
were obtained  for Uranus ($\tau_a/\tau_R=0.96$) and for Neptune
($\tau_a/\tau_R=1.35$), and comparison of these data shows that
the amount of aerosol in Neptune is larger than in Uranus(fig.4).

\section{Conclusion}
In this paper we determined the values of aerosol to gas ratio of
optical depth component for Uranus ($\tau_a/\tau_R=0.96$) and for
Neptune ($\tau_a/\tau_R=1.35$), and spectral dependence of
absorbing to scattering ratio $\tau_\kappa/\tau_S$.

We confirmed that ignoring of real temperature profile leads to 50
$\%$ errors in determination of ($\tau_a/\tau_R$).

Real spectral values of single scattering albedo corrected for
Raman scattering, were obtained for spectral region 350-450 nm.

\section{Acknowledgements}

I thank A.V.Morozhenko and A.P.Vidmachenko for good advices and
useful discussion of obtained results.

\begin {references}
\reference M.J.S. Belton, L.Wallace, M.J. Price: 1973, "\apj",
184, N3, p.143-146; \reference W.D. Cochran, L.M. Trafton: 1978,
"\apj", 219, N1, p.756-762; \reference L.Delbouille, G. Roland,
L.Neven : 1973, "Liege:Univ.press"; \reference M.Dementiev: 1992,
"Kinematika i physika nebesnyh tel", 8, N2, p.25-35; \reference
E.Karkoschka: 1994, "Icarus", 111, N3, p.967-982;\reference G.F.
Lindal, J.R. Lyons, D.N. Sweetnam, et al. : 1987,
"J.Geophys.Res.", A92, N3, p.14987-15001;\reference G.F. Lindal,
J.R. Lyons, D.N. Sweetnam, et al. : 1990, "J.Geophys.Res.", 17,
N10, p.1733-1736; \reference A.V.Morozhenko: 1997, "Kinematics and
Physics of Celestial bodies", 13, N4, p.20-29; \reference
O.V.Morozhenko: 2004, "Naukova Dumka", p.206; \reference
A.V.Morozhenko: 2006, "Kinematika i physika nebesnyh tel", 22, N2,
p.138-153; \reference A.V.Morozhenko and N.Kostogryz : 2005,
"Kinematika i physika nebesnyh tel", 21, N2, p.114-120; \reference
J.B.Pollack, K.Rages, K.H.Baines, et al.: 1986, "Icarus", 65,
N2/3, p.442-466;\reference L.Wallace: 1972, "\apj", 176, N1,
p.249-257;
\end {references}
\end{document}